\documentclass[ap, journal]{copernicus}
\begin{document}


\title{Diffuse Synchrotron Emission from Galactic Cosmic Ray Electrons}

\Author[1]{Giuseppe}{Di Bernardo}
\Author[2]{Dario}{Grasso}
\Author[3]{Carmelo}{Evoli}
\Author[4,5]{Daniele}{Gaggero}
\affil[1]{MPI f\"ur Astrophysik, Karl-Schwarzschild-Strasse 1, D-85740 Garching, Germany}
\affil[2]{Istituto Nazionale di Fisica Nucleare, Sezione di Pisa, Largo B. Pontecorvo, I-56127, Pisa, Italy}
\affil[3]{II. Instit\"ut f\"ur Theoretische Physik, Universit\"at Hamburg, Luruper Chaussee 149, D-22761 Hamburg, Germany}
\affil[4]{SISSA, Via Bonomea 265, I-34136 Trieste, Italy}
\affil[5]{ INFN, sezione di Trieste, via Valerio 2, I-34127, Trieste, Italy}
\runningtitle{Diffuse Synchrotron Emission from Galactic Cosmic Ray Electrons}
\runningauthor{Giuseppe Di Bernardo}
\correspondence{Giuseppe Di Bernardo (bernardo@mpa-garching.mpg.de)}

\received{}
\pubdiscuss{} 
\revised{}
\accepted{}
\published{}


\firstpage{1}
\maketitle
\begin{abstract}
Synchrotron diffuse radiation (SDR) emission is one of the major Galactic components, in the 100 MHz up to 100 GHz frequency range. Its 
spectrum and sky map provide valuable measure of the galactic cosmic ray electrons (GCRE) in the relevant energy range, as well as of the 
strength and structure of the Galactic magnetic fields (GMF), both regular and random ones. This emission is an astrophysical sky foreground 
for the study of the Cosmic Microwave Background (CMB), and the extragalactic microwave measurements, and it needs to be modelled as better 
as possible.
In this regard, in order to get an accurate description of the SDR in the Galaxy, we use - for the first time in this context - 3-dimensional 
GCRE models obtained by running the \textsc{Dragon} code. This allows us to account for a realistic spiral arm pattern of the source 
distribution, demanded to get a self-consistent treatment of all relevant energy losses influencing the final synchrotron spectrum. 
\end{abstract}

\introduction  
Deflection of ultra-high energy cosmic rays (UHECR), rotation measure, synchrotron radiation, and polarized dust are just a small sample of different methods of observation able to investigate the galactic magnetized interstellar medium (ISM).  
In this frame, cosmic rays (CRs) are, doubtless, a unique probe of the ISM properties.
Thanks to a set of successful experiments such as Fermi-LAT, PAMELA, AMS-02, the last few years have witnessed an incredible progress in the knowledge of electron and positron Galactic CRs, over a wide range of energy, from $\mathcal{
O}(\textrm{TeV})$ down to tens of \textrm{MeV}. 
Unfortunately, solar modulation complicates matters, since the CR spectra observed on Earth are - for $E \lesssim 20$ \textrm{GeV} - completely reshaped with respect to their local interstellar spectra (LIS). 
\par 
Relativistic cosmic ray electrons and positrons (CRE), 
spiralling around the interstellar magnetic field lines, 
are at the origin of the radio diffuse emission from the Milky Way. 
For magnetic field intensity of $\mathcal{O}(\mu \textrm{G})$, like in the case of our Galaxy, and for electrons/positrons of [\textrm{GeV$\div$TeV}] energies, the synchrotron emission falls in the [\textrm{MHz$\div$GHz}] range. Indeed, the SDR offers valuable complementary checks of the low energy spectrum, and in general of the spatial distribution of CRs in the Galaxy.
Hence, a parallel study of radio emission, together with CR measurements, can put better constraints on all the interstellar medium (ISM) components involved \citep{2011A&A...534A..54S}.
 The interpretation of those measurements requires a proper
modelling of injection, propagation and losses in the Galaxy.  \par 
Moreover, the presence in the [$20 \div 200$] GHz range of several astrophysical sky signal components - with similar intensities and some spatial correlation - makes the extraction of the CMB a complex task. In order to achieve sufficient accuracy on the cosmological signal the component separation needs to take advantage of the knowledge on the properties of diffuse Galactic emission.  
\par
We plan to accomplish the aforementioned study by running the \textsc{Dragon} code in its $3$-dimensional version. Indeed, this is
well suited to model the CRE propagation, when accounting for a realistic spiral arm distribution of astrophysical sources, gas distributions, magnetic fields models and different position-dependent models for diffusion in the parallel and perpendicular directions with respect to the GMF.  

\begin{table}[t]
\caption{The CRE models considered in the present analysis. The reported values of $\gamma_{\textrm{inj}}$($e^{-}$) refer to energies below/above $4$ \textrm{GeV}.}
\begin{tabular}{l c c c r}
\tophline 
Model & $\delta$ & $v_{A}(\textrm{km s}^{-1})$ & $\eta$ & $\gamma_{\textrm{inj}}$($e^{-}$) \\
\middlehline
\textsf{KRA} & 0.5 & 15 & -0.4 & 1.6/2.5 \\  
\middlehline
\textsf{KOL} & 0.33 & 35 & 1.0 & 1.6/2.5 \\ 
\bottomhline
\end{tabular}
\label{table:tabella}
\belowtable{} 
\end{table}
\section{Objectives and method}
In the present Section, we outline the guidelines of the \textit{multi-wavelength} analysis we have performed, in order to model the CRE spectra consistently with the diffuse synchrotron emission of the Galaxy.
One of our main aims has been:
\begin{enumerate}
\item To explore the physical properties - injection and propagation - of the local interstellar spectrum (LIS) of CRE, in the realm of low energies ($\lesssim 7$ \textrm{GeV}), by combining the latest $e^{-}$ and $e^{+}$ measurements with the diffuse Galaxy radio emission, between $10$ \textrm{MHz} and few \textrm{GHz}. Below that energy, we modelled the LIS of $e^+ + e^-$ on the basis of the observed synchrotron spectrum of the Galaxy, which is unaffected by propagation in the heliosphere (see also e.g., \citet{2013MNRAS.436.2127O});
\item In parallel to that, the current study has pushed us to give an important constrain on the vertical scale height of the diffusion region in the Galaxy, by looking simultaneously at the radio spectrum, the latitude profile of the synchrotron emission, and the positron fraction at energies below $\sim 5$ GeV.
\end{enumerate}
\par 
The structure of the GMF is still not well understood. Generally,   
a realistic, and accurate description of the synchrotron emission, as well as of its angular distribution, requires to take in account two main components for the GMF: the regular and turbulent ones. 
Regarding the ordered one, here we rely on a recent model, based on a wide and updated compilation
of Faraday rotation measurements \citep{2011ApJ...738..192P}. It consists of two different components: a \textit{disc} field, with a magnitude in the vicinity of the solar system, $B_{0}$, taken to be $2~\mu$\textrm{G}, and a toroidal \textit{halo} field, with a thickness of $\sim[0.2 \div 0.4]$ \textrm{kpc}, and extended - above, and below - out of the galactic plane (GP) for $[1 \div 2]$ \textrm{kpc}\footnote{We have included the regular field in order to make our model compatible with the current information. However, in our analysis, we have checked that only the halo component ($B_{\textrm{halo}} = 4~\mu$\textrm{G}) for the regular GMF plays, albeit marginal, a role.}.
A new, and much-improved model for the regular GMF has been recently brought to the attention of the community \citep{2012ApJ...757...14J}. 
Besides a disc field and an extended halo field, the peculiarity of this GMF model is its \textrm{X}-shape in the $r-z$ directions. 
The main implications based on such a ordered GMF model will be addressed in our future work.  
\par 
Instead, for what concerns the random component - actually, the main responsible for the diffusion of charged particles in the ISM -  there is still a poor knowledge about its geometrical structure. As in \citet{2004ApJ...610..820H}, we have assumed it to fill a thick disk, modelled with an exponential vertical profile, and an effective scale-height $z_{h}$, accordingly to the equation: 
\begin{equation}
\label{eq:turbolento}
B_{\textrm{ran}}(z) = B_{\textrm{ran}}(0)\exp{(-z/z_{h})}.
\end{equation} 
\begin{figure}[t]
\includegraphics[width = 6.0 cm] {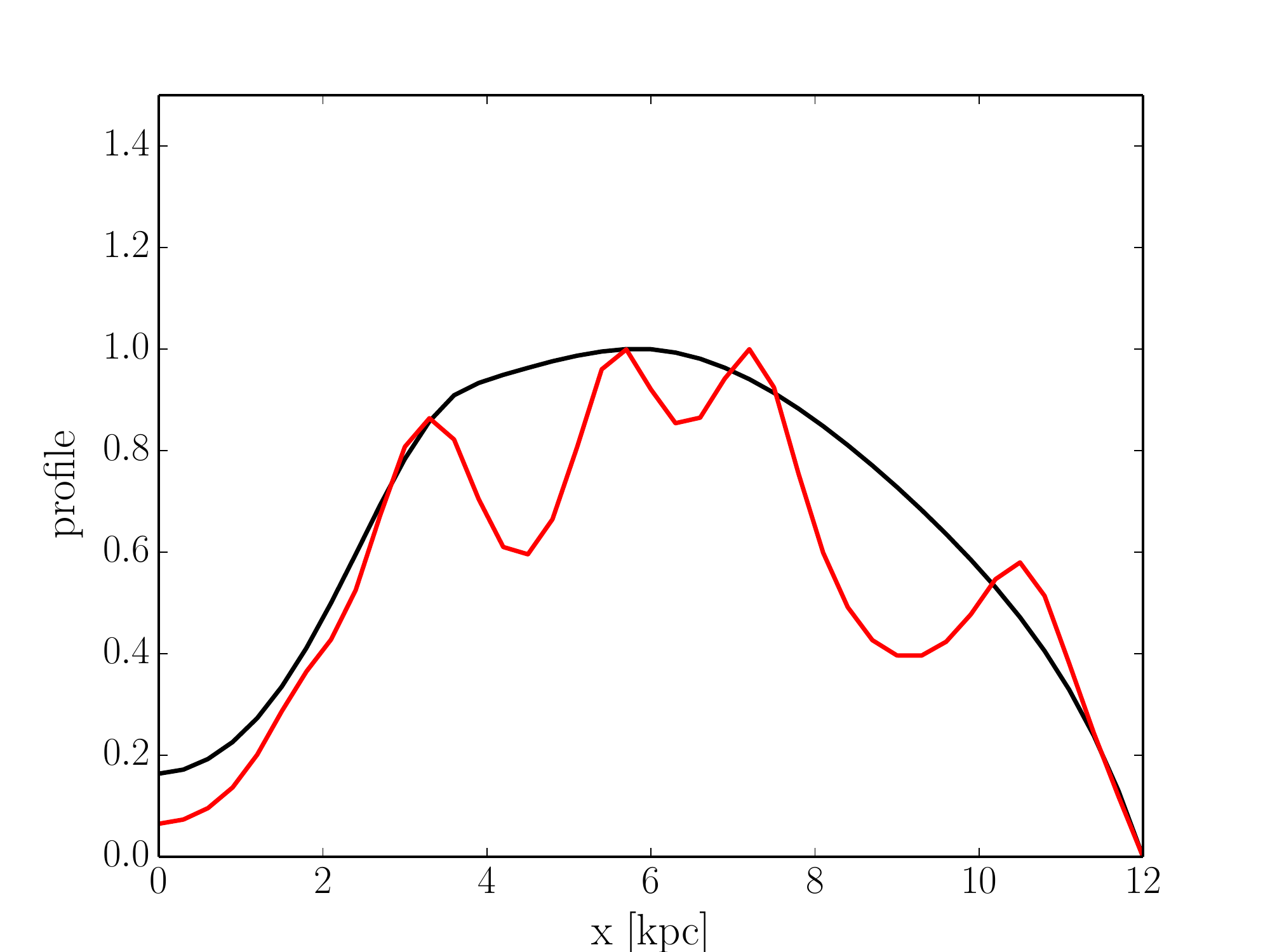}
\caption[]{Normalized electron density profile along the radial direction. The black line corresponds to a 2-dimensional, smooth CR. 
The red one corresponds to a 3-dimensional spiral arm pattern of sources.}
\label{fig:2Dvs3D}
\end{figure}
\par 
For the aforementioned purposes, we first run \textsc{Dragon}, a new numerical package aiming to solve the diffusion equation of CRs in the Galaxy environment \citep{Evoli:2008, Gaggero:2013}.
\par 
Conversely to a too simplified assumption of a diffusion spatially uniform in the thick disc, here in this contribution we account for a possible spatial dependence of the diffusion coefficient, 
\begin{equation}
D(\rho,R,z) = D_{0}\beta^{\eta}f(z)\biggr(\frac{\rho}{\rho_{0}}\biggl)^{\delta},
\end{equation}
$\rho$ being the rigidity of the particle, $\beta$ the particle speed in units of speed of light $c$, and $f(z)$ indicates the spatial dependence of the diffusion coefficient. As predicted by the \textit{quasi-linear theory} (QLT), that should be related to the fluctuating magnetic field, and hence as $D(z)^{-1} \propto B_{\rm ran}(z)\propto \exp{(-z/z_{h})}$.
\par 
In this paper, for sake of simplicity, only two representative classes of propagation regimes have been taken in consideration: the \textsf{KRA} (Kraichnan), and the \textsf{KOL} (Kolmogorov). The main parameter characterizing those two models are reported in Table \ref{table:tabella}. 
\par 
For each of them, we varied the scale-height of the diffusive halo in the range $z_{h} = [1 \div 16]$ kpc, and the main diffusive parameters were determined in order to
to minimize the combined $\chi^2$ against the boron-to-carbon ratio and the proton observed spectra. 
\par 
Finally, at high energies ($\gtrsim 7$ \textrm{GeV}), unlike our previous results presented in \citet{DiBernardo:2012}, here we fix the spectral index, and the normalization of the injection spectrum of the primary electrons and of the extra-component by tuning our models against the new data, as recently released by PAMELA and AMS-02 collaborations, respectively, rather than on $e^{-} + e^{+}$ spectrum measured by Fermi-LAT \citep{2014PhRvD..89h3007G}. With reference to the Figure \ref{fig:2Dvs3D}, we want to make it clear that the assumption of a simple power-law (PL) distribution for energetic electrons and positrons, whose sources are smoothly distributed in the entire Galactic disc, leads to large overestimation of their energy densities in comparison with the values deduced when a 3-dimensional spiral arm distributions are used. In our opinion, the impact of this more realistic modelling of the particle distribution on the final synchrotron spectral maps is a crucial issue (see also the comparison between the Figures \ref{fig:spectralmap_2d} \& \ref{fig:spectralmap_3d}).  
\section{The synchrotron emission of the Galaxy}
It is known that the synchrotron intensity depends on the spatial, and energetic distribution of CRE density, $n_{e}$, and the strength of the magnetic field ($B_{\perp}$), perpendicular to the line of sight (LOS) to the observer.
Once the CRE densities are computed by \textsc{Dragon} - at all points of the computational grid - we use \textsc{Gammasky} to get the emissivities (i.e. power per unit volume per unit frequency per unit solid angle), for the regular and random fields, according to the standard formalism \citep{2011hea..book.....L}, after having integrated over the particle energy.
\par
The emissivity (in \textrm{erg} \textrm{s}$^{-1}$ \textrm{Hz}$^{-1}$) - in an uniform magnetic field - is partially linearly polarized\footnote{for a monochromatic and isotropic distribution of CRE.}, and usually subdivided in two components, $\epsilon_{\parallel,\perp}$, respectively parallel and perpendicular to $B_{\perp} \equiv B(\vec{r})\sin\alpha$, with $\alpha$ the angle between the direction of the magnetic field and the LOS. The polarization formulation will be used in our future work. Here, we show results based only on the total intensity, given by
\begin{equation}
\label{eq:regular}
\epsilon(\nu, \vec{r}) = \sqrt{3}\frac{e^{3}}{mc^{2}}B_{\perp}(\vec{r})F(x);
\end{equation}
where $x = \nu/\nu_{c}^{\textrm{reg}}$, being $\nu_{c}^{\textrm{reg}} = (3/4\pi)(e/mc)B_{\perp}\gamma^{2}$ the critical synchrotron  frequency, $\gamma$ the particle (electron or positron) Lorentz factor, and $F(x)$ is defined in terms of Bessel functions.
In the case of a randomly oriented magnetic field, the expected isotropic emissivity is computed according to \citet{1988ApJ...334L...5G}.
\begin{figure}[t]
\includegraphics[width=6.5 cm]{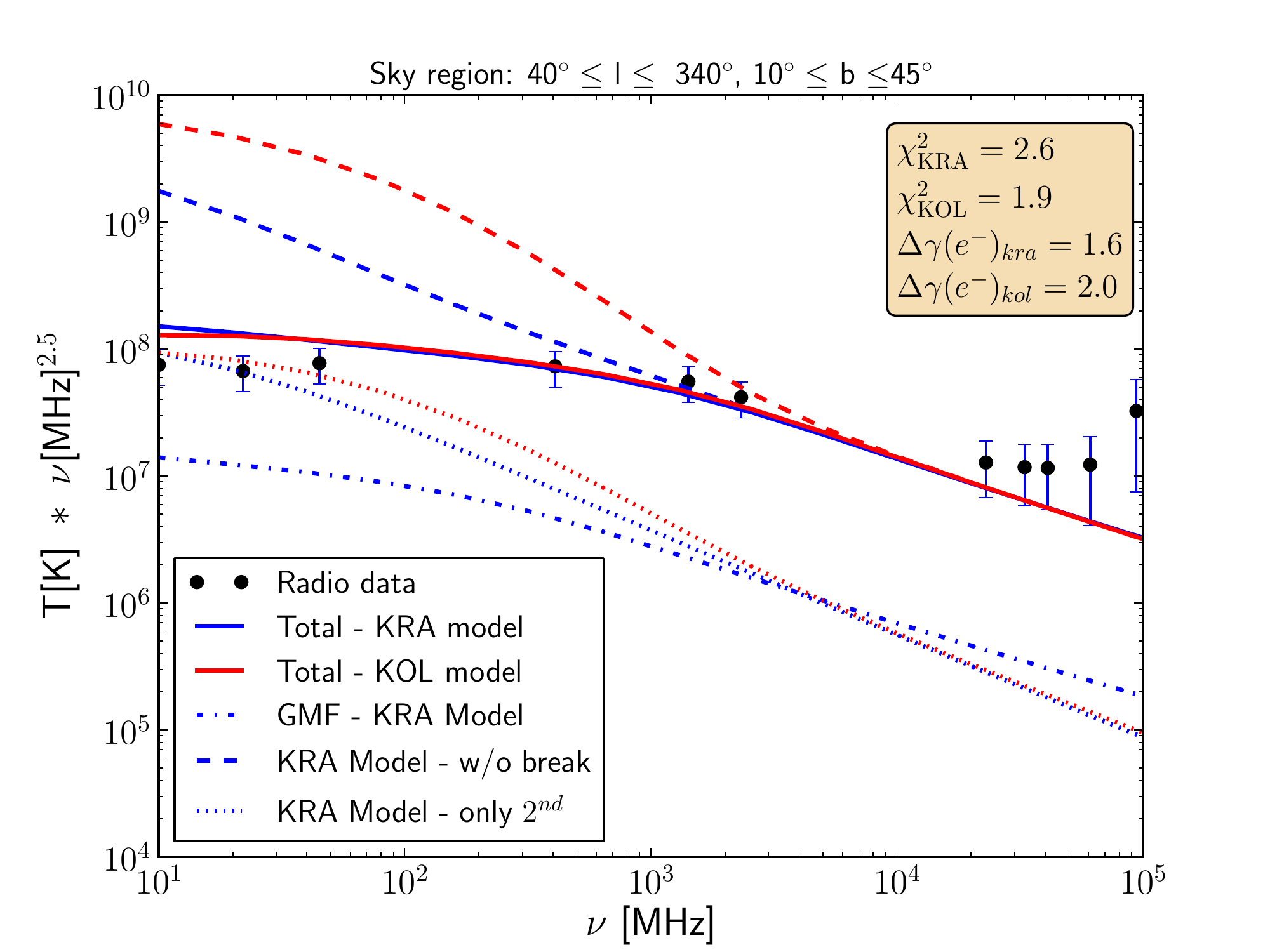}
\caption[Synchrotron spectrum]{The average synchrotron spectra, for $z_{h} = 4$ \textrm{kpc}. We show the spectra obtained with (solid lines) and without (dashed lines) the spectral break in the $e^{-}$ source spectra. Dotted lines are the contribution of secondary $e^{-}$ source spectra. The contribution of the regular GMF is shown as the dot-dashed line. The  The normalization required for the random component field strength is $B_{\textrm{ran}}(0) = 7.6~\mu\textrm{G}$.}
\label{fig:spettro}
\end{figure}

\begin{figure}[t]
\centering
\includegraphics[width=6. cm]{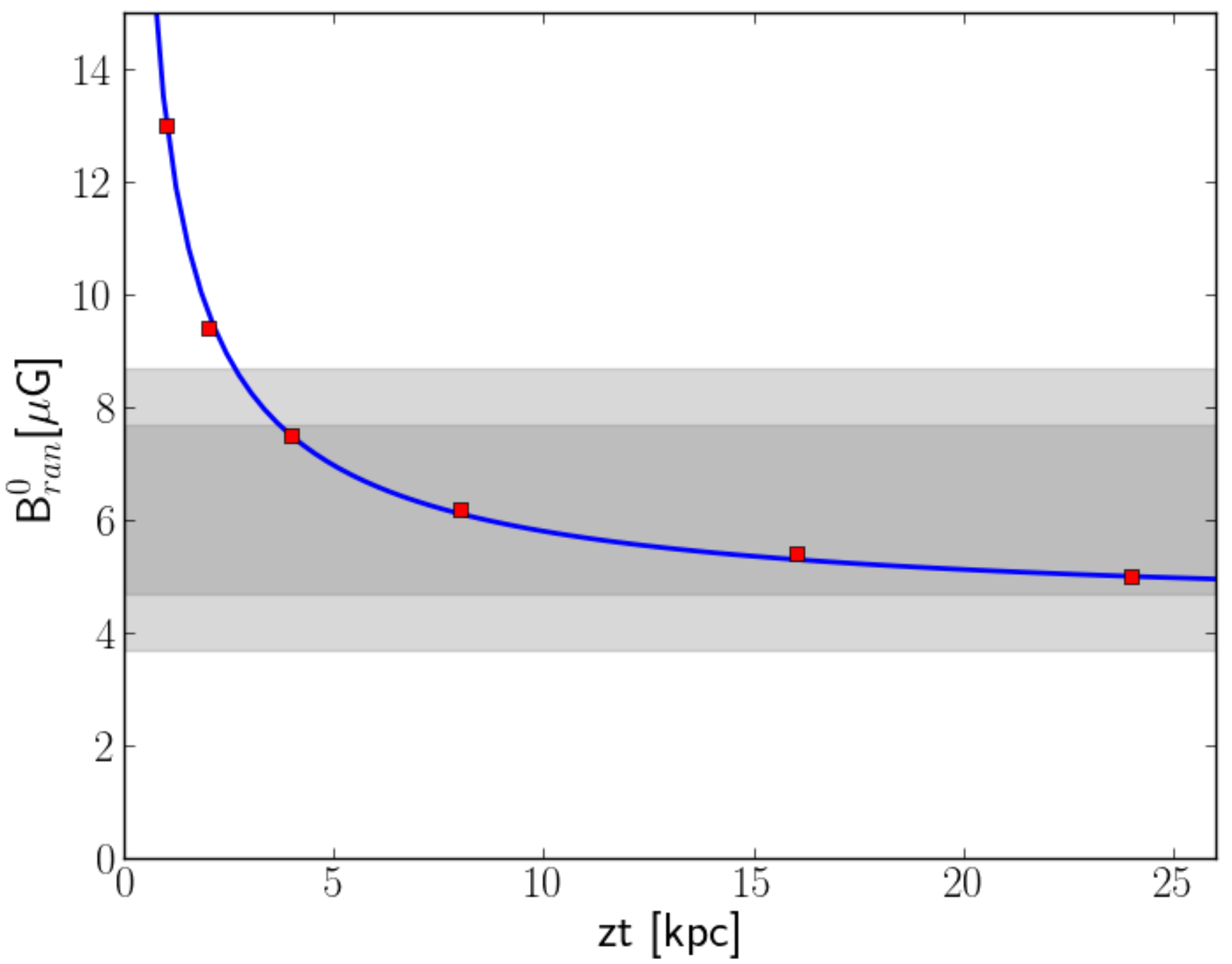}
\caption[normalization]{Normalization of random GMF \textit{vs} the vertical scale height. The $3(5)~\sigma$ allowed by RM are represented in grey (light grey) bands. The red squares are the values used in our \textsf{KRA} models in order to reproduce the observed spectrum at $408$ \textrm{MHz}.}
\label{fig:normalization}
\end{figure} 
\subsection{The total synchrotron intensity}
Given a GMF model, and for the representative CRE density models aforementioned, the next step is to compute the Galactic synchrotron spectrum. We take care of correctly reproducing the observed $408$ {\rm MHz} radio synchrotron radiation as in \citet{1982A&AS...47....1H}, by tuning - time to time - the normalization value for the turbulent component of the GMF (see the Eq. \ref{eq:turbolento}). 
We get the sky maps in \textsc{Gammasky} by integrating the Galactic emissivity along the LOS,
\begin{equation}
\label{eq:intensity}
I(\nu) = \int_{l.o.s}\epsilon(\nu, \vec{r})ds,
\end{equation}
where $\epsilon(\nu, \vec{r})$ is the total emissivity given by the Eqs. \eqref{eq:regular}. In Figure \ref{fig:spettro}
we refer to the observed \textit{brightness} temperature (in \textrm{K}), defined as \citep{1986rpa..book.....R} 
\begin{equation}
T(\nu) = \frac{c^{2}I(\nu)}{2k_{B}\nu^2}.
\end{equation}
\par   
The sky maps are subdivided into equal area pixels following the \textsf{HEALPix}\footnote{http://healpix.jpl.nasa.gov/} pixelization scheme of \citet{2005ApJ...622..759G}. We average the flux over the sky regions $40^\circ < l < 340^\circ,  10^\circ < b < 45^\circ, -45^\circ < b < -10^\circ$, where $l$ and $b$ are Galactic longitude
and latitude respectively.
We restrict the analysis to the regions out of the Galactic plane - but avoiding the polar regions - being the contamination from point-like and local extended sources expected to be the smallest. In addition to that, we avoid absorption effect at radio frequencies, and free-free emission at higher frequencies \citep{1986rpa..book.....R}. Therefore, the observed Galactic diffuse emission in the radio band is, almost entirely, due to the synchrotron radiation of CRE moving errantly in the GMF. 
\par 
From tens of \textrm{MHz} to $23$ GHz (and up to 94 \textrm{GHz} for \textsc{Wmap}), in such sky region we directly compare our simulated models with the synchrotron spectra measured by a wide set of radio surveys at $\textbf{22}$, $\textbf{45}$, $\textbf{408}$, $\textbf{1420}$, $\textbf{2326}$ \textrm{MHz} as well as \textsc{Wmap} satellite data at $\textbf{23}$, $\textbf{33}$, $\textbf{41}$, $\textbf{61}$ and $\textbf{94}$ \textrm{GHz}.
\par 
With reference to the Figure \ref{fig:spettro}, it is immediate to realize that the radio data ($\lesssim \mathcal{O}(100)$ \textrm{MHz}) are clearly incompatible with a single PL electron spectrum, suitable to fit the CRE data. Rather, we find that introducing - below a few \textrm{GeV} - either a break or an exponential \textit{infrared} (IR) cut-off in the population of primary $e^-$, that helps us in providing a very good description of the radio data. 
\par
As immediate consequence of that, the total electron flux (at $E \lesssim 4$ \textrm{GeV}), and hence the radio spectrum below
$100$ {\rm MHz}, are dominated by secondary particles, which are produced in nuclear collision with the nuclei of the gas present in the ISM, offering thus a direct probe of the interstellar proton spectrum. 
In this regard, we found that once the low energy $e^-$ source spectrum is tuned to reproduce the observed $e^+$ spectrum, only models featuring low re-acceleration can reproduce the observed $e^+$ spectrum and fraction, in total agreement with what found in \citet{2011A&A...534A..54S, 2013MNRAS.436.2127O}.
\subsection{The magnetic halo height}
The vertical - perpendicular to the Galactic plane - size of the CR diffusion region represents, undoubtedly, a \textit{cornerstone} in modern Astro-particle physics. The accurate knowledge of it goes beyond the target of conventional CR astrophysics; it is also worth for \textit{Dark Matter} (DM) indirect search, since that the local flux of DM decay, and annihilation products are expected to depend significantly on such physical length.
\begin{figure}[t]
\centering
\includegraphics[width=7.0 cm]{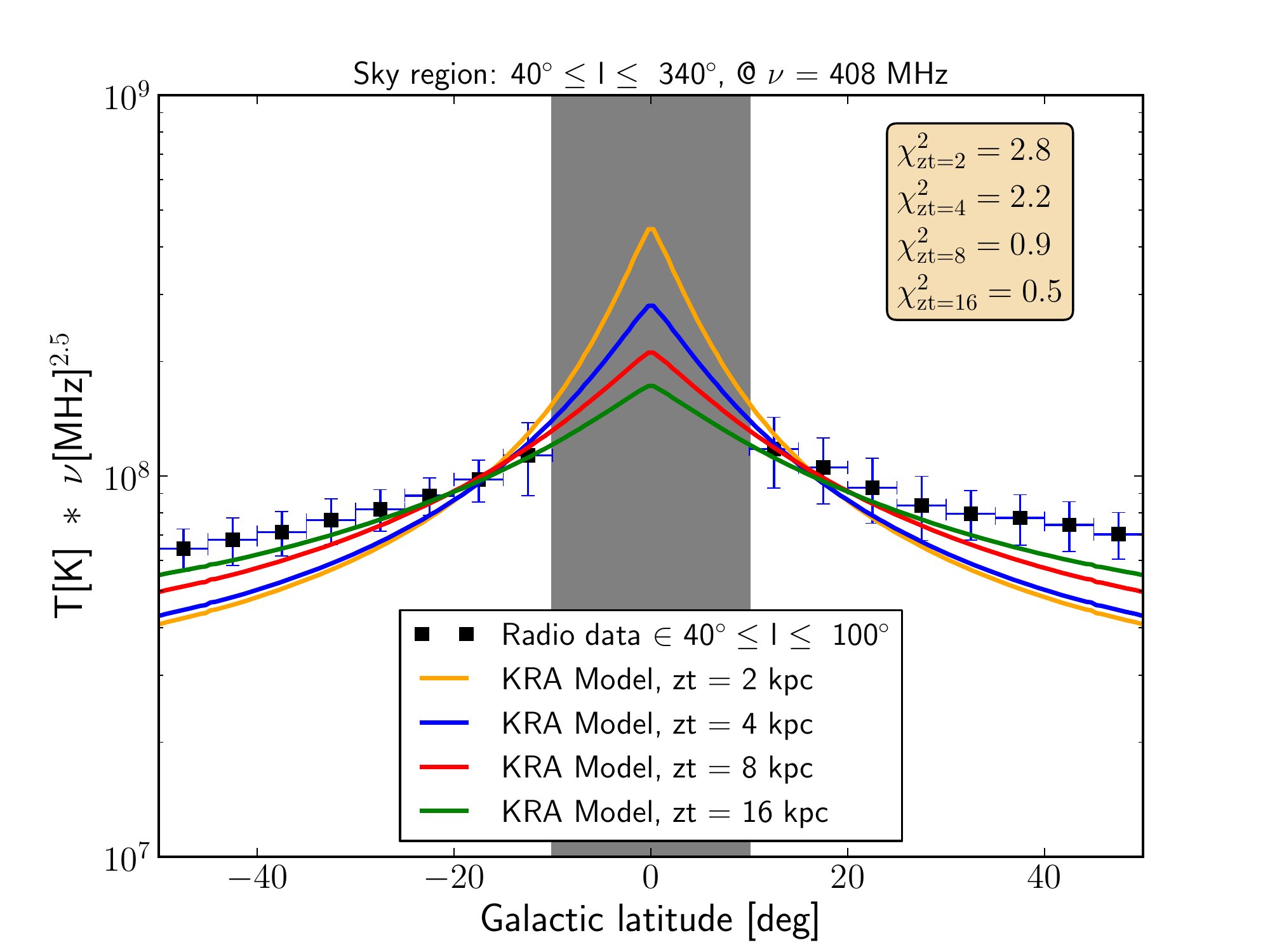}
\caption[Latitude profile]{The latitude profile for the synchrotron emission at $408$ \textrm{MHz}, at different magnetic halo height.}
\label{fig:profile}
\end{figure}  
\par 
So far, the diffusive vertical boundary has been constrained purely on the basis of CR radionuclide, $^{10}Be/^{9}Be$ ratio most commonly. However, this method is marked by several flawless, due to the severe uncertainties connected to local distribution of sources, gas, and especially by the solar modulation.
\par
To the contrary, the synchrotron emissivity of the Galaxy offers a much
more genuine probe of the scale height $z_{h}$. In that regards, we
may notice that, when a realistic vertical distribution is adopted
for the radiation interstellar field (ISRF) and for the GMF, energy
losses in the $\mathcal{O}$(GeV) energy range - hence in the  radio energy band - do not affect significantly the CRE
vertical distribution, determined predominantly by the diffusion and
therefore coincident with that of CR nuclei (see Figure 5 in \citep{DiBernardo:2012}).
\par 
Our first argument aiming to constrain the value of $z{_h}$ is the following one: for a given propagation set up, the synchrotron flux depends, from Eq. \eqref{eq:intensity}, only on the random field normalization $B_{\textrm{ran}}(z = 0)$, dominant respect to the regular one,  and on $z_{h}$, the scale-height of the diffusion region ($I(\nu,\vec{r}) \propto \int B(\vec{r})n_{e}(\vec{r})ds$). As it is possible to appreciate in the Figure \ref{fig:normalization}, the fit of radio data suggests a tight relation $B_{\rm ran}^{2}(z = 0)\propto z_{h}^{-1}$.
\par 
Secondly, we compare the observed latitude profile of the synchrotron emission at $408$ {\rm MHz} to that calculated for the \textsf{KRA} set up, setting different values for $z_{h}$. For each $z_{h}$ we tune the value of $B_{\textrm{ran}}(z = 0)$ so that the average spectrum in these regions is
reproduced (Figure \ref{fig:profile}). Low values of $z_{h}$ are disfavoured: a $\chi^2$
analysis showed that $z_{h} \leq 2$ kpc are excluded at $3\sigma$ level.

\section{Future directions and conclusions}
To fully observe and understand the GMF, there is a long way to go. 
With our analysis, we exploited the SDR as a way to measure the low energy LIS spectrum of CRE. For the first time, we have placed a constraint on the CR diffusive halo scale height, based on the comparison of the computed synchrotron emission intensity with radio observations.
Moreover, we stress out that - for the first time in this framework - our modelling of the SDR emission accounts for the presence of the $e^\pm$ \emph{charge-symmetric extra-component},
required not only to consistently model PAMELA and AMS-02 high energy data, but also to correctly estimate the $e^-$ source spectrum from CRs and radio data. In our opinion, the combination of high precision CRE data, and current radio observations, can be a viable method to disentangle the contribution of the extra-component to the total synchrotron spectrum. 
\par 
One of the greatest challenges of observing the Cosmic Microwave Background (CMB) in the [$20\div200$] \textrm{GHz} range resides in the separation between the CMB and the superimposed foreground emission: free-free, synchrotron, thermal dust. 
We pointed out that transport of charged relativistic particles, and magnetic fields models should be studied simultaneously, 
because both have influence on the synchrotron modelling.
Synchrotron spectra may reveal signatures of spatially inhomogeneous particle source distributions and magnetic fields. In our treatment here, instead of a smooth CR distribution invariant for rotations about the Galactic disc axis, we rather calculate self-consistently the synchrotron maps emitted by electrons whose spectral density is inhomogeneous, due to all the relevant energy losses sustained while traversing regions with different distributed gas, magnetic and radiation fields. 
\par 
Finally, the frequency range of \textsc{Gammasky} synchrotron simulations, from $\mathcal{O}(10)$ \textrm{MHz} to $\mathcal{O}(100)$ \textrm{GHz}, 
covers radio telescopes such as Planck, LOFAR, and SKA. We have been developed \textsc{Gammasky} with the aim to support the scientific exploitation of the data provided by these experiments.
\begin{figure}[t]
\includegraphics[width=7. cm]{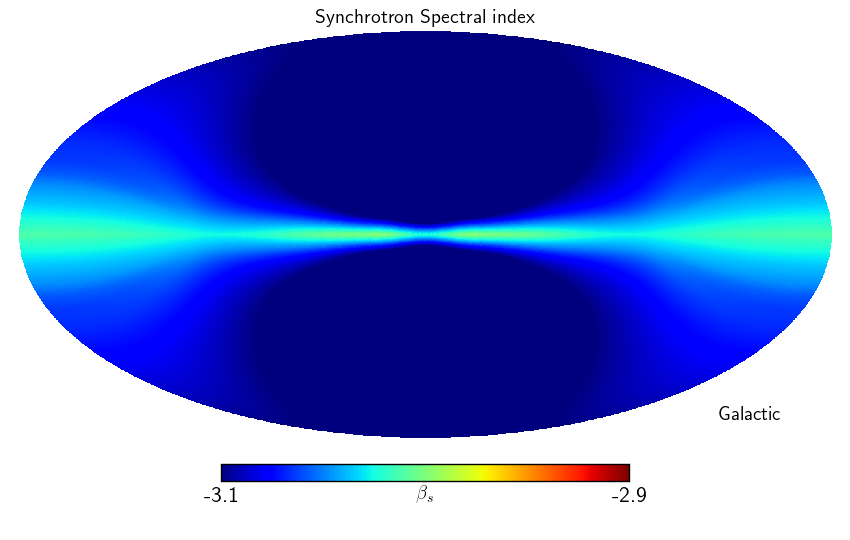}
\caption[]{Spectral index map between $408$ MHz and $23$ GHz, with the assumption of a smooth $2$D CR sources. Given the total intensity $I(\nu)\propto (\nu/\nu_{0})^{\beta_{s}}$, the map has been computed according the standard formalism: $\beta_{s} = 0.248\log(I_{23}/I_{408})$.}
\label{fig:spectralmap_2d}
\end{figure}
\begin{figure}[t]
\includegraphics[width=7.cm]{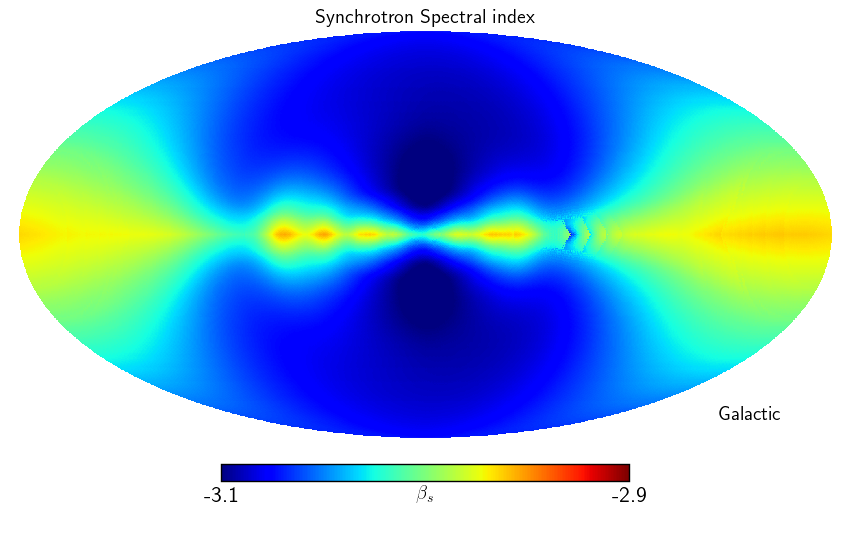}
\caption{As in Figure \ref{fig:spectralmap_2d}, but with the important assumption of a $3$D spiral arm structure.}
\label{fig:spectralmap_3d}
\end{figure} 







\bibliographystyle{copernicus}
\bibliography{Bibliography.bib}
\end{document}